# Multifunctional tapered fiber-based micro-waveguide for optical ultrasound microsensors

**MENGYUE ZHANG,[1] CHANGHUI LI[1,2,*]**

[1] *Department of Biomedical Engineering, College of Future Technology, Peking University, Beijing, China*

[2] *National Biomedical Imaging Center, Peking University, Beijing, China*

*\*chli@pku.edu.cn*

Various optical ultrasound microsensors have been developed with size ranging from tens to hundreds of micrometers. However, it becomes challenging to further minimize these sensors' size. In this work, we proposed a method that use a tapered fiber-based micro-waveguide (TFMW) attaching to the optical microsensor to bypass this challenge. The TFMW not only serves as the waveguide to transport ultrasound, but can also deliver light and actively excite ultrasound. In this study, we proposed the design and analyzed its performance using theoretical analysis and simulation.

## 1. BACKGROUND

To achieve high-resolution ultrasound (US) and photoacoustic (PA) imaging of micro-structures, ultrasound detectors with wide bandwidth, high sensitivity and extremely small size are desired. However, transducers based on the piezoelectric (PZT) material have limitations of relatively narrow bandwidth and quickly reduction in sensitivity as the detection size becomes small[1]. The optical ultrasound sensing based on optical resonance is emerging as a promising technology for miniaturized, high-sensitivity detection, which maintains high sensitivity even when reduced to tens of micrometers[2].

Several types of optical microresonators have been developed. Micro-ring whispering-gallery modes (WGM) resonators with high quality-factor, typically fabricated on silicon or polymer substrates, have sizes of the order of tens of micrometers[3-5]. Another extensively explored architecture is the fiber Fabry–Pérot(FP) interferometric cavity, formed by two parallel semi-reflective mirrors on fiber tips[6-8]. Besides, the Fiber Bragg Grating (FBG) sensor relies on a small resonant cavity formed by introducing a π-phase shift in the Bragg grating to achieve strong acousto-optic coupling, enabling sensitive detection with footprints below 100 micrometers[9-11].

Despite these advances, due to technical difficulty and reduction in the quality factor, it becomes very challenging to further reduce the size of the optical resonator-based ultrasound detector down to ten-micrometer scale, while maintaining high performance. Furthermore, for on-chip micro-sensors, they generally require extra space (millimeter scale) for affiliated structures to deliver probing light, hindering the miniaturization of the sensor's volume[3, 4, 12]. Besides the micro-resonator structure itself, to protect the sensitive sensor from contamination or corrosion in liquid environment, extra coating is also often required, leading to larger size[13, 14].

To address these challenges, in this work, we propose the tapered fiber-based micro-waveguide (TFMW) to decouple the resonator sensing cell from the detection tip, realizing micrometer-range detection size while maintaining superior optical resonator performance. Using optical fiber as the ultrasound waveguide have been studied longtime ago [15, 16]. For instance, Prof. Zou reported using a commercial optical fiber of 125 μm as the waveguide to deliver ultrasound below 3.375 MHz[17]. However, most of these fiber US waveguides are not suitable to propagate high-frequency signals (tens to hundreds of MHz) due to their large diameter. Here, we explored using the tapered fiber with diameters down to 10-μm scale, enabling to deliver much higher frequency ultrasound. In addition, the same optical fiber can also deliver light to locally excite ultrasound itself. We theoretically studied properties of this micro-waveguide via simulation and analysis, demonstrating its capability of effectively collecting high-frequency ultrasound waves with high fidelity, as well as uniquely exciting ultrasound via photoacoustic effect.

## 2. CONCEPT OF THE MICRO-WAVEGUIDE

The tapered optical fiber is commonly used, for example, the coupling fiber for WGM resonator has a typical diameter of several micrometers. The concept of TFMW for US sensing is illustrated in Figure 1(a). The micro-waveguide delivers ultrasound from its far end remotely to optical sensor bodies, including FBG, FP and WGM sensors. The TFMW can be either fused/glued to the sensor body (Fig. 1(b)) or as a seamless natural extension of the sensor structure itself (Fig. 1(c)). Since for a typical optical fiber made of fused silica, the ultrasound attenuation is very low[18], the length of waveguide can be as long as several centimeters or longer.

The TFMW decouples the detection tip with the sensor body, allowing detection of ultrasound in liquid while the sensor body remains in the air. It is an important advantage of TFMW since immersing the optical sensor into the liquid environment often requires additional protective methods. For instance, the WGM sensor generally requires special coating to prevent itself from contamination. The temperature in liquid also affects the performance of the sensitive optical sensor. Furthermore, owing to the chemical and physical durability of optical fibers (such as those made by fused silica), the probe can operate in harsh environments, such as those with high acidity, high alkalinity, or high temperature.

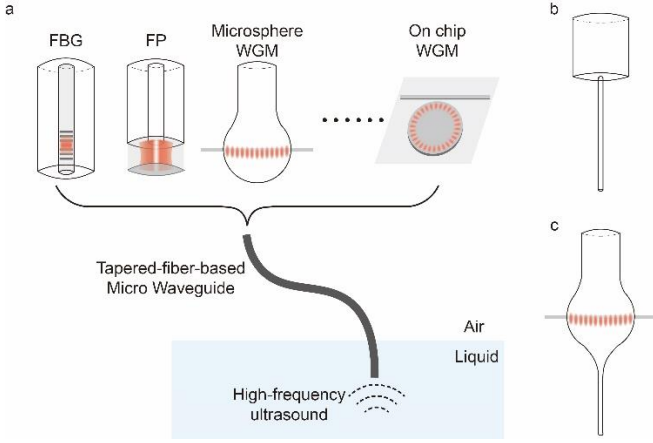

Fig. 1 Schematic of the TFMW concept. The micrometer scale tapered optical fiber delivers ultrasound from liquid regime to optical sensors in the air.

## 3. MODELING OF WAVE INTERACTIONS WITH TFMW

Since the TFMW is made of optical fiber, it is inherently capable of conducting both light and ultrasound, forming a dual-modal waveguide. This unique characteristic makes TFMW have multifunctional potential. In the following, to evaluate the performance of the TFMW as a dual-modal waveguide for both ultrasound and light, we chose the material of the tapered fiber to be the common fused silica, which has typical properties listed in Table 1.

Table 1. Physical parameters of fused silica

| Density | Young's modulus | Poisson's ratio | Longitudinal wave velocity | Refractive index | |
|---|---|---|---|---|---|
| | | | | $n_{core}$ | $n_{cladding}$ |
| 2203 kg/m³ | 73.1 GPa | 0.17 | 5847 m/s | 1.4504 | 1.4437 |

### 3.1 Ultrasound interaction with TFMW

The effective propagation of ultrasound with high fidelity is crucial for ultrasound detection, which requires low dispersion and low attenuation. The tapered fiber can be approximately regarded as an infinite long solid rod. To effectively suppress dispersion, it is ideal to only allow longitudinal wave at its fundamental single mode to propagate, which requires $d \ll \lambda$, where $d$ is the diameter of the TFMW and $\lambda$ is the acoustic wavelength inside the TFMW. As previously explored, when $d < 0.2\lambda$, the TFMW can't support high-order modes, which suppresses the severe signal distortion caused by mode dispersion[15, 19]. According to this relationship, the original 125-μm fiber only allows propagation of ultrasound with a frequency below 10 MHz without dispersion, which is not suitable for high-frequency waveguide.

In the simulation, we set the diameter of the waveguide to be 10 μm, which can theoretically support up to 120 MHz. Limited to the coumputing resource, we set the center frequency of the acoustic wave to be 30 MHz, corresponding to wavelength of about 200 μm in the fiber. This satisfies the condition for suppressing high-order guided modes.

The acoustic coupling characteristics of the waveguide system were studied using finite-element simulations (COMSOL Multiphysics, Version 6.2, COMSOL AB, Sweden). Figure 2(a) shows the simulation setup. To conveniently simulate plane waves with different incident angles, the bottom of the simulation domain was set to be a semi sphere. Plane-wave radiation boundaries are applied to the surface of the liquid area, which act as absorbing layers, effectively mimicking an unbounded medium while also enabling plane-wave excitation. The ultrasound source is a pulsed plane wave, defined as follows,

$$p(\pmb{x},t) = p_0 G(\pmb{x},t)\sin\left(2\pi f_0\left(t - \frac{\pmb{x}\cdot\pmb{C}}{c|\pmb{e_k}|}\right)\right), \quad (1)$$
$$G(\pmb{x},t) = e^{-\frac{\pi^2 f_0^2}{2\ln 2}\left(t - \frac{\pmb{x}\cdot\pmb{e_k}}{c|\pmb{e_k}|}\right)^2}$$

where $p_0$ is the pressure amplitude of the wave, $G(\pmb{x},t)$ is the Gaussian envelope, $f_0$ is the carrier signal frequency, and $\pmb{e_k}$ is the wave direction vector. Plane waves incident from different directions to the waveguide were simulated.

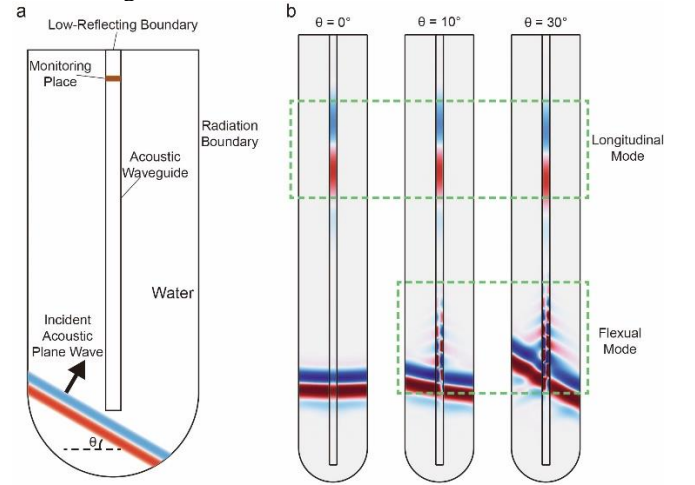

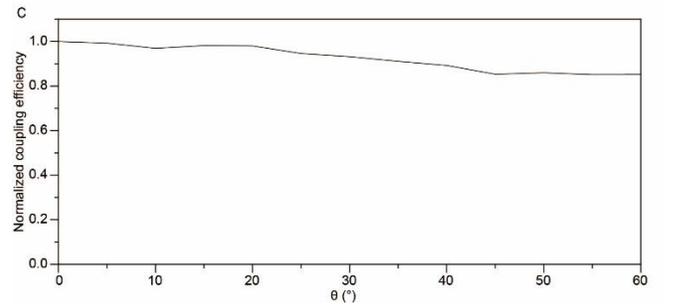

Fig. 2 Simulation of the ultrasound interaction with TFMW. (a) Simulation settings; (b) Snapshots of the distribution of ultrasound pressure waves in TFMW at a time after incidence of pulsed plane wave at three incident angles; (c) Dependence of normalized coupling efficiency on incident angles.

Figure 2(b) shows the acoustic pressure fields inside the waveguide after propagation over a certain distance for 0, 10, and 30 degrees of incident angles, respectively. When the plane wave is normally incident from below, only the longitudinal mode with a phase velocity of near 6000 m/s is excited. In contrast, oblique incidence excites both longitudinal and flexural modes[15]. Fundamental mode of longitudinal wave propagates in tapered fiber very fast, as marked in the upper dashed box in Figure 2(b). In the frequency range of 0 MHz to 100 MHz, the phase velocity of the flexural modes varies from approximately 0 m/s to near 3000 m/s. The much slower speed and stronger dispersion result in a delayed and progressively broadened signal train relative to the longitudinal mode, marked in the lower dashed box in Figure 2(b). As the TFMW long enough, the effect of delayed flexural waves can be eliminated in time domain for sensing pulsed ultrasound. For instance, with a 2-cm long TFMW, even with 3000 m/s flexural wave, there is a 3.3 μs delay in arrival time relative to the faster longitudinal wave, which can be feasibly separated.

To investigate coupling efficiency, the energy coupled into the waveguide's longitudinal mode is evaluated from the peak integrated mechanical energy flux over a cross-section traversed by the mode('Monitoring Place' in Figure 2(a)). The incident energy is taken as the peak integrated acoustic intensity over the same area for the incident plane wave. The ratio of the two indicates that, for normally incident sound waves, the energy conversion efficiency is approximately 17% for acoustic signal coupling from the liquid into the waveguide's longitudinal mode. Then, normalized to the normal incidence case, Figure 2(c) presents the power coupling efficiency dependence on the plane-wave incident angle.

It is worthy to note that the extremely small fiber tip of the TFMW makes it feasible to approach the target to very close distance, i.e., in micrometer scale. Therefore, the TFMW could have a large solid angle to collect US signal. Besides, closer to the target significantly reduced the medium attenuation for US, especially for high-frequency US waves.

### 3.2 Light interaction with TFMW

Owing to its inherent property, the TFMW can also deliver light and emit light out of the tip. At micrometers distance from the tip, the emitted light from the tip is not seriously diverged, which can be used for photoacoustic excitation. Interestingly, we found different treatments applied to the fiber tip can yield different excitation effects. Once we shrink the tip region, as in Figure 3(a), it enhances optical confinement near the tip and produces a smaller output spot size, enabling micrometer-scale optical resolution for photoacoustic excitation.

We simulated light propagation and emitting out of the tapered fiber tips, as illustrated in Figure 3(a). Fiber was immersed in water, and perfectly matched layers were applied around the simulation domain to emulate an unbounded space. Based on the 10-μm diameter, a single-mode optical field (λ = 532 nm) was launched through a port at the left side of the fiber.

Figure 3(b) presents the electrical field distributions for three representative tip diameters, with tip radius of 1.5 μm, 2 μm and 5 μm, respectively. It can be seen that electrical field was confined in a short distance before diverging. Therefore, if TFMW is approaching an optically absorbing target at micrometer distance, it can excite photoacoustic signal with high resolution. Figure 3(c) shows the full width at half maximum (FWHM) of the output beam at different distances from the tip. According to the results, thinner tips produce smaller focal spots, reaching sub-micrometer scales; however, these tightly focused beams diverge rapidly. In contrast, thicker tips generate larger spot sizes but maintain a slow varying beam width over longer distances. This flexibility allows the design to accommodate different application needs: thinner tips for higher lateral resolution, and thicker tips when larger depth of field is needed.

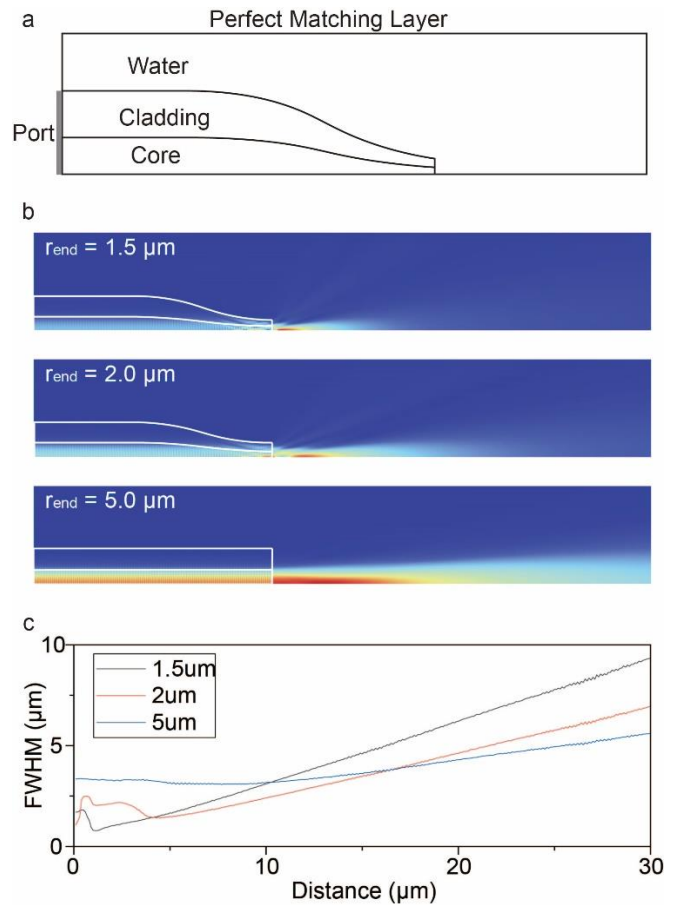

Fig. 3 Simulation of the pattern of light emitting out of TFMW tip. (a) Simulation settings; (b) Simulation results at three tip diameters; (c) Variation of the emitted beam spot size with propagation distance after emission.

Shrinking the tip region alters the fiber structure. However, since the major TFMW is unchanged, it won't affect the overall ultrasound propagation characteristics of the waveguide. Our study shows that it only slightly influences the coupling efficiency.

## 4. MULTIFUNCTIONAL TFMW

Based on its ultrasound and light co-propagating characteristic, TFMW offers a variety of potential application scenarios, as illustrated in Figure 4. Figure 4(a) shows its role as a micro acoustic sensor for receiving external acoustic waves. Figure 4(b) demonstrates its implementation in ultrasonic pulse/echo detection, in which an absorptive coating is applied to the tip, and the pulsed light delivered via TFMW excites the absorbing coating to generate a pulsed ultrasound via photoacoustic effect. Thus, the TFMW produces a broadband microscale US source suitable for microscopic ultrasound imaging or sensing. Figure 4(c) illustrates its application in photoacoustic imaging or sensing, where light emits out of the TFMW tip to excite photoacoustic signal of the target.

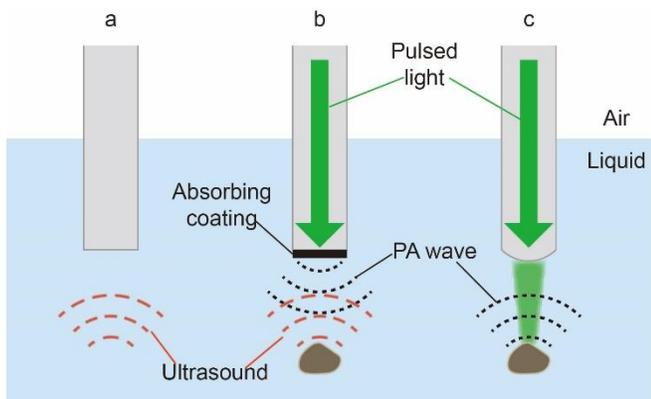

Fig. 4 Three typical functions of TFMW. (a) Pure ultrasound sensor; (b) Ultrasound pulse/echo mode via photoacoustic effect; (c) Photoacoustic sensor.

## 5. DISCUSSION

In summary, we propose a versatile dual-modal micro-waveguide based on tapered optical fiber, TFMW, which enables various optical ultrasound sensors to achieve US sensing size down to 10-μm scale. Besides, TFMW decouples the sensing tip with the sensor body, allowing sensing ultrasound in liquids while maintaining the sensor body in air. Theoretical analysis and simulation demonstrate the high performance of TFMW in US sensing and photoacoustic excitation.

Based on these unique properties and characteristics of TFMW, there are multiple functions it can provide, including US sensing, US pulse/echo, and photoacoustic excitation. We investigate TFMW's acoustic and optical behavior through theoretical simulations. This approach is compatible with multiple detection modalities and offers high flexibility. It has the potential of expanding the applicability of microcavity sensors without increasing fabrication complexity, enabling microcavities of diverse materials, geometries, and sizes to be used in extreme-environment sensing or in scenarios requiring a micro-detection system, such as endoscopic imaging.

Future works need to manufacture such micro-waveguides that attached to optical micro-resonators. Besides technics, wave propagation and interaction will become more complicated than the simulation. For instance, once we connect TFMW with the sensor body, at the jointing regime of these two components (TFMW and sensor body), both acoustic impedance and morphology might vary, these will disturb the US wave, whose influence and strategies to optimize need further investigation case by case.

**Funding.** This research was supported by the following grants: the National Key R&D Program of China (No. 2023YFC2411700); the Beijing Natural Science Foundation, China (No. 7232177).
**Disclosures**. The authors declare no conflicts of interest.